\documentclass[graphics,twocolumn, usenatbib]{mn2e}

\usepackage{epsfig} 
\usepackage{graphicx} 
\usepackage{color}

\usepackage{aas_macros} 
\usepackage{amssymb}

\newcommand{\lya}{Lyman-$\alpha$~}
\newcommand{\gad}{\textsc{Gadget-2~}}
\newcommand{\enzo}{\textsc{Enzo~}}
\newcommand{\cosmos}{\textsc{Cosmos~}}
\newcommand{\enzoamr}{\textsc{Enzo(AMR)~}}
\newcommand{\enzost}{\textsc{Enzo(static)~}}

\def\etal{{\it et al.}~}

\title[Numerical simulations of the \lya forest]{Numerical Simulations
of the \lya forest -- A comparison of \textsc{Gadget-2} and \textsc{Enzo} }

\author[J.A. Regan \etal] 
{John A. Regan$^1$\thanks{E-mail:regan@ast.cam.ac.uk}, Martin G. Haehnelt$^{1,2}$ \& Matteo Viel$^{1,3}$  \\ \\
$^1$ Institute of Astronomy, Madingley Road, Cambridge CB3 0HA \\
$^2$Kavli Institute for   Theoretical Physics,   Kohn Hall, UCSB, Santa Barbara CA 93106\\
$^3$ INAF - Osservatoria Astronomico di Trieste, Via G.B. Tiepolo 11, I-34131 Trieste, Italy \\}

\begin{document}

%Make the Title
\maketitle

%%%%%%%%%%%%%%%%%%%%%%%%%%%%%%%%%%%%%%%%%%%%%%%%%%%
%Abstract time
\begin{abstract}
We compare simulations of the \lya forest performed with two different
hydrodynamical codes,  \gad and \textsc{Enzo}. A comparison  of the dark matter
power spectrum for simulations run with identical  initial conditions
show differences of  1-3\% at the  scales relevant for quantitative
studies of the  \lya forest. This allows a meaningful comparison  of
the effect of the  different implementations of the
hydrodynamic part of the two codes. Using the same  cooling and
heating algorithm in both codes   the differences in the  temperature
and the density probability distribution function are of the order of
10 \%. The differences are comparable  to the
effects  of boxsize and resolution on these statistics. 
When self-converged results for each code are taken into account 
the differences in the flux power spectrum -- the statistics
most widely used  for estimating  the matter power spectrum and
cosmological parameters from \lya forest  data --  are  about 5\%. This
is again comparable to the effects of boxsize  and
resolution. Numerical uncertainties  due to a particular
implementation  of solving the hydrodynamic or gravitational equations
appear therefore to contribute only moderately to the error budget
in estimates of  the flux power  spectrum from numerical simulations. We
further find that the differences  in the flux power spectrum for \enzo
simulations run with and without adaptive mesh refinement are also of 
order 5\% or smaller. The latter require 10 times less CPU time making 
the CPU  time requirement
similar to that of a version of \gad that is optimised  for \lya
forest simulations.
\end{abstract}
%%%%%%%%%%%%%%%%%%%%%%%%%%%%%%%%%%%%%%%%%%%%%%%%%%%
%keywords time
%%%%%%%%%%%%%%%%%%%%%%%%%%%%%%%%%%%%%%%%%%%%%%%%%%%
%Introduction time

\section{Introduction}

There is  now a well established  paradigm for the origin  of the \lya
forest,  the  ubiquitous  absorption  lines due  to neutral hydrogen
in the  spectra of  high-redshift quasars.  The absorption blue-wards
of 1216 $\AA$ is predominantly due to density fluctuations in the
intervening warm  ($\sim 10^4$ K) photoionized inter-galactic medium 
(IGM)  on scales larger
than  the Jeans  length  of the  gas  (see \cite{Rauch_1998} for a
review).  Numerical simulations  were instrumental in  establishing
the new paradigm in the 1990s (\cite{Cen_1994}, \cite{Zhang_1995},
\cite{Hernquist_1996}, \cite{Theuns_1998a}, \cite{Zhang_1997}). 
%\cite{Zhang_1998}   
The  \lya forest and the associated  metal
absorption probe the thermal and ionization history of the
IGM  as well as the interplay of galaxies and
the IGM from which they are formed. More recently  the \lya forest has
also  been  established  as a means   of  quantitative measurement of
the underlying matter distribution and  thus a variety of cosmological
parameters ({\it e.g.} \cite{Croft_1998}, \cite{Croft_2002a},
\cite{Viel_2004}, \cite{Seljak_2005a}, \cite{Viel_2006b},
\cite{Seljak_2006}; Viel, Haehnelt \& Lewis (2006)).   Numerical simulations thereby play a crucial
role in  inferring the linear matter power spectrum and other derived
quantities from the \lya forest data.  With increasing sample   sizes
statistical errors of measurements of the flux  distributions  have
reached the percent level and the  error budget is  dominated  by
systematic uncertainties  (Viel, Haehnelt \& Springel (2004),
\cite{McDonald_2005a} \nocite{McDonald_2005b}).  Uncertainties due to
numerical simulations  contribute significantly to the error budget
and  the accuracy with  which  the  flux distribution  for   given
input  physics  and cosmological parameters can be simulated  has
become important.  Most studies so far have used convergence tests to  assess
uncertainties due to the numerical simulations and direct 
comparisons of cosmological hydrodynamical simulation performed with
different codes have been rare.
The  differences between hydro-dynamical  simulations of galaxy
clusters with a wide range of different codes/methods have been
studied in the Santa Barbara cluster project (Frenk et
al. 1999)\nocite{Frenk_1999}.   Recently \cite{O'Shea_2005} performed
a  comparison  between the grid based adaptive mesh  refinement (AMR) code
\textsc{Enzo}\footnote{http://cosmos.ucsd.edu/enzo/} and the smoothed particle hydrodynamics 
(SPH) code \textsc{Gadget-2}\footnote{http://www.mpa-garching.mpg.de/gadget/}.  
However, little  has   been  done  in
this  respect  for hydrodynamical simulations of the \lya  forest data
(see \cite{Theuns_1998b} for a notable exception of a comparison
between two SPH codes) 
%\cite{Zhan_2005} for a comparison of
%pseudo-hydro techniques and \cite{Machacek_2000} for a comparison of
%cosmological models).  
Some comparisons of hydrodynamical simulations with approximate
simulations of the \lya forest data have been  performed   by    
McDonald  et al.  (2005),\cite{Zhan_2005}  and Viel, Haehnelt \& Springel  (2006).
We  present here a comparison of hydrodynamical simulations of the
\lya  forest with \enzo and \gad which concentrates on the statistical
properties of the flux distribution. \\ We are therefore mostly
interested in  properties of the moderate to low over-density  gas  which
is  responsible  for \lya forest absorption.  Of particular interest
is the probability distribution  of the gas density, the temperature,
the resulting flux distribution and the  flux power spectrum.  A major difference
between grid-based and SPH codes is
their treatment of shocks and their effects on the temperature
distribution. We will also examine these differences.\\  \indent The
plan of  the  paper is as follows.  In \S\ref{Codes} we describe  the
\enzo code and the \gad code and the different  ways  in  which  the
codes  solve  the gravitational  and hydrodynamics  equations. In
\S\ref{Sims}  we describe  the simulation set used in the
comparisons. In \S\ref{comparison} we investigate how physical
properties  of both codes  compare,  in  particular the  gas
distribution and the 1-D flux power spectrum. Finally in \S\ref{time}
we will  look at the performance of each code in terms of CPU time
consumption.

%Section describing the two codes Enzo & Gadget

\section{Hydrodynamical simulations of the \lya forest}
\label{Codes}

\subsection{Grid-based simulations {\it vs.} SPH simulations}  

The physical state of the gas responsible for  the  \lya forest is largely
governed by the competing processes of photoionisation and adiabatic
cooling due to expansion. The low over-density gas obeys a tight  temperature - density
relation, (e.g. \cite{Katz_1996}, \cite{Hui_1997} ) which  can be
approximated by

\begin{equation}
T = T_0 \Big( {\rho_b \over \overline \rho_b} \Big)^{\gamma - 1}
\end{equation}
where $\rho_b $ and  $\overline \rho_b$ are the baryon density and
mean baryon density respectively, $T$ is the temperature, $T_0 $ and
$1<\gamma<1.6$ are parameters which depend on the reionisation history
model and  the  spectral shape of the UV background. Typical
temperatures of the photoionized IGM are in the range  10000-20000
K. The optical depth for  \lya absorption is proportional to
the neutral hydrogen density, \citep{GunnPeterson_1965}, which, since
the gas is in photoionisation equilibrium, is proportional to the
square of the density  times the recombination rate,

\begin{equation}
\tau \propto \Big({\rho_b \over \overline \rho_b}\Big)^{2} T^{-0.7} =
A \Big({\rho_b \over \overline \rho_b}\Big)^{\beta}
\end{equation}
where $\beta = 2.7 - 0.7 \gamma$. The factor A depends on the
redshift, baryon density, temperature at the mean density, Hubble
constant and the photoionisation rate. The optical depth is   a
faithful tracer of the matter distribution  on scales larger than the
Jeans length of the photoionized IGM.  Even though the density field
is only mildly non-linear on the relevant scales the thermal  effects, the
peculiar velocities and the non-linear relation  between flux and
optical depth make hydrodynamical  simulations mandatory for  accurate
predictions of the statistical properties of the flux  distribution in
the \lya  forest. Cosmological hydrodynamical simulations come in  two
basic flavours, SPH (Lucy 1977, Gingold \& Monaghan 1977) \nocite{Lucy_1977} 
\nocite{Gingold_Monaghan_1977}
and grid based codes.  SPH codes like
\gad use particles to represent the baryonic fluid. SPH is a
Lagrangian method and hence  the  resolution is concentrated in
regions of high density. Grid based codes  like \textsc{Enzo}, use a grid of
cells to represent the gas properties.  One of the options in \enzo  is
an adaptive refinement of the grid where the  grid resolution
is increased in regions of  high density resulting in a large dynamic
range. Both codes have been used to study the \lya forest (e.g Viel,
Haehnelt \& Springel (2004), 
\cite{Tytler_2004}, \cite{Bolton_2005}, \cite{Jena_2005}, \cite{McDonald_2005b}).  
SPH simulations of the \lya forest in particular have been very successful
but  in principle one may think that a grid based code could offer
better  resolution of the low density IGM than SPH codes because
in SPH  simulations the resolution in low density regions decreases
when the gravitational clustering of the matter distribution becomes
non-linear  and high density regions start to form. Grid-based codes
could also  offer  a more accurate  treatment of shocks  which are
relevant for the  thermal state of the IGM. We will now  briefly
describe the methods implemented  in  \enzo and \gad to solve the
gravitational   and hydrodynamical equations.

%%%%%%%%%%%%%%%%%%%%%%%%%%%%%%%%%%%%%%%%%%%%%%%%%%%
%%%%%%%%%%%%%%%%%%%%TABLE 1%%%%%%%%%%%%%%%%%%%%%%%%
\begin{table}
\begin{tabular}{ |l |l |c}
%        \hline   &  \textbf{\em  \gad   Simulations}  &   \\  \hline
\hline \hline 
\textbf{\em  Particle number/Rootgridsize} &  \textbf{\em Boxsize ($h^{-1}$ Mpc)} &\textbf{\em SDR}\\ 
\hline 
\hline 
\textbf{\em \enzo AMR Grid} \\
$50^3$ & 15.0\ \ \ \ \ \ 30.0\ \ \ \ \ \ 60.0 & 800\\
$100^3$ & 15.0\ \ \ \ \ \ 30.0\ \ \ \ \ \ 60.0 & 1600\\ 
$200^3$ & 15.0\ \ \ \ \ \ 30.0\ \ \ \ \ \ 60.0 & 3200\\ 
\hline 
\textbf{\em  \enzo Static Grid} \\ 
$50^3$ & 15.0\ \ \ \ \ \ 30.0\ \ \ \ \ \ 60.0 & 50\\ 
$100^3$ & 15.0\ \ \ \ \ \ 30.0\ \ \ \ \ \ 60.0 & 100\\ 
$200^3$ & 15.0\ \ \ \ \ \ 30.0\ \ \ \ \ \ 60.0 & 200\\ 
$400^3$ & 15.0\ \ \ \ \ \ 30.0\ \ \ \ \ \ 60.0 & 400\\ 
\hline  
\textbf{\em  \gad }\\ 
$50^3$ & 15.0\ \ \ \ \ \ 30.0\ \ \ \ \ \ 60.0  & 800\\ 
$100^3$ &15.0\  \ \ \ \ \  30.0\ \ \ \ \  \ 60.0 & 1600\\ 
$200^3$ & 15.0\ \ \ \ \  \ 30.0\ \ \ \ \ \ 60.0 & 3200\\ 
\hline  
\textbf{\em  \gad (star formation)}\\ 
$400^3$ & 15.0\ \ \ \ \  \ 30.0\ \ \ \ \ \ 60.0 & 6400\\ 
\hline 
\hline
\end{tabular}
\caption{Spatial parameters and spatial dynamic range (SDR) for the
\enzo AMR simulations, the \enzo static grid simulations 
and the \gad simulations with and without star formation.  The particle numbers
for \gad and the rootgridsize for \enzo is shown in the left hand
column.  In the \gad simulations the number of DM particles and SPH
particles is equal and for the \enzo simulations the number of DM 
particles is equal to the rootgridsize.}
\label{Gadget_sims}
\end{table}
%%%%%%%%%%%%%%%%%%%%%%%%%%%%%%%%%%%%%%%%%%%%%%%%%%%%%%%%%%%%%%%%%%%%%

\subsection{\enzo} 

\enzo is a Eulerian adaptive mesh refinement code 
originally developed  by Greg  Bryan  and Mike  Norman at  the
University of  Illinois (Bryan \& Norman 1995b, Bryan \& Norman 1997, 
Norman \& Bryan 1999, O'Shea et al. 2004).
\nocite{Bryan_1995b} \nocite{OShea_2004} \nocite{Bryan_1997} \nocite{Norman_1999}
The hydrodynamics solver employs the  Piecewise Parabolic  method
combined with  a non-linear  Riemann solver  for  shock  capturing.
The  Eulerian AMR  scheme  was  first developed   by
\cite{Berger_1984} and later refined by \cite{Berger_1989} 
to  solve   the  hydrodynamical equations for an ideal gas.  
\nocite{Bryan_1995b} Bryan \& Norman 1997  adopted such  a   scheme  for
cosmological   simulations.    The gravity solver in \enzo uses a
N-Body    particle    mesh    technique (Efstathiou et al. 1985, 
Hockney \& Eastwood 1988). 
\nocite{Efstathiou_1985} \nocite{Hockney_1988}\\
\indent We have used the publicly available version of \enzo (\enzo-1.0.1) which we have 
modified to use the \gad equilibrium chemistry solver as discussed in \S\ref{enzo_hydro}.

\subsubsection{\enzo gravity solver} 

The gravity solver in \enzo employs an adaptive particle-mesh
algorithm.   The potential is solved  on a  periodic root grid  using
Fast Fourier Transforms.  In  order  to  accurately  account  for  the
sub grids  a multi-grid relaxation technique  is used (see e.g. Norman \& Bryan 1999). 
The force resolution is typically twice as coarse  as the grid resolution.

%%%%%%%%%%%%%%%%%%%%%%%%%%%%%%%%%%%%%%%%%%%%%%%%%%%%%%%%%%%%%%%%%%%%%%%%%%%
%%%%%%%%%%%%%%%%%FIGURE%%%%%%%%%%%%%%%%%%%%%%%%%%%%%%%%%%%%%%%%%%%%%%%
\begin{figure*}
  \centering           \begin{minipage}{175mm}          \begin{center}
  \psfig{figure=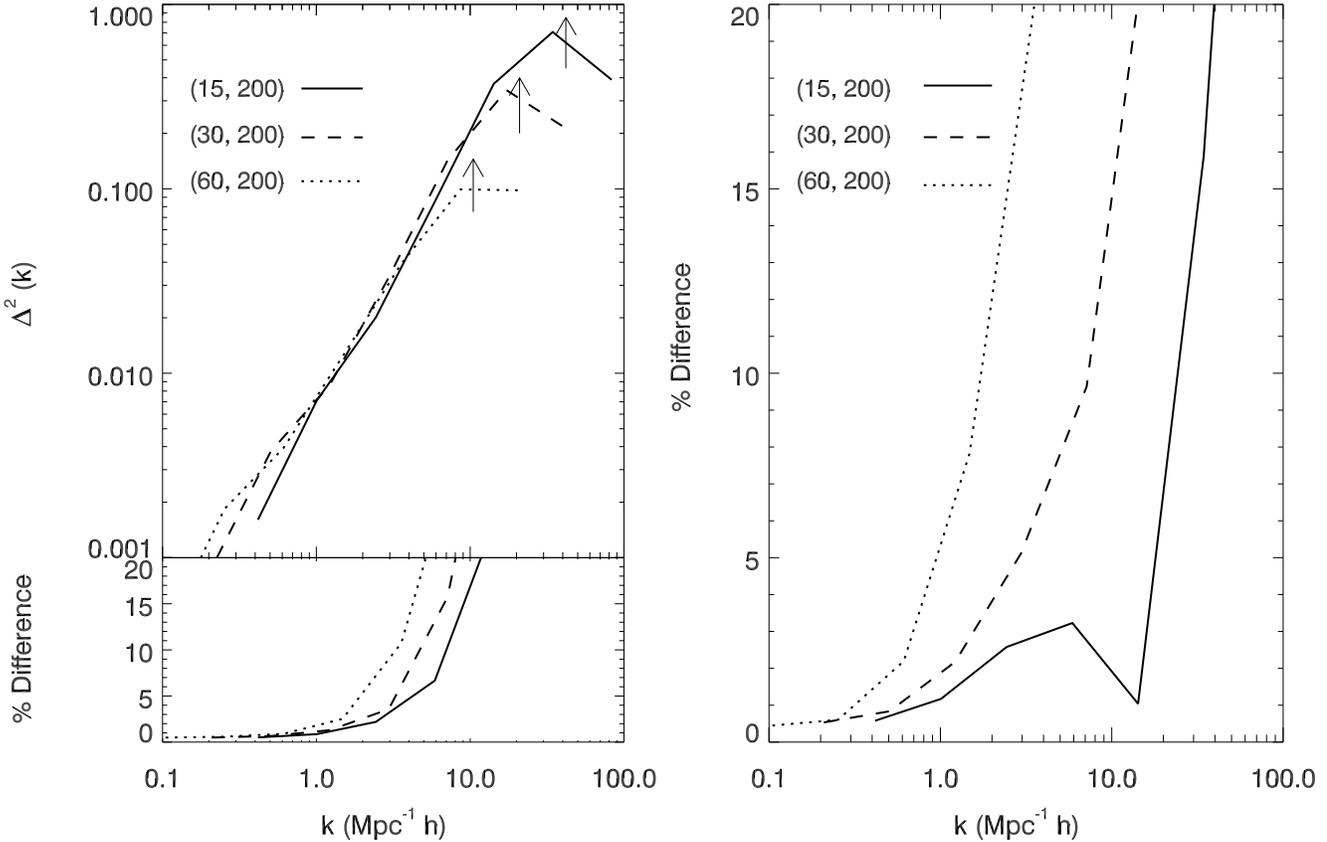,width=1.0\textwidth}
  \caption{\label{DMPS}  {\it Left Panel:} {\it Top:} The DM power spectrum of the
    \enzo AMR simulations with a rootgridsize of $200^3$. The arrows indicate
    the Nyquist frequency for the rootgrid for the 60 $h^{-1}$ Mpc , 30 $h^{-1}$ Mpc
    and 15 $h^{-1}$ Mpc boxes (comoving) from left to right. {\it Bottom:} 
    The fractional  difference between the Enzo 
    simulations with and without AMR [(\enzoamr - \enzost) /
      \enzoamr].
    {\it Right Panel:} The fractional difference between the  DM power spectrum of
  simulations with \gad and \enzoamr [(\gad - \enzoamr) / \gad]. All  results
  are for $z=3.0$. }

  \end{center} \end{minipage}
\end{figure*} 

%%%%%%%%%%%%%%%%%%%%%%%%%%%%%%%%%%%%%%%%%%%%%%%%%%%%%%%%%%%%%%%%%%%%%%%%%%%%

\subsubsection{\enzo hydrodynamics solver} 
\label{enzo_hydro}
\enzo    uses     the    Piecewise    parabolic     method    (PPM),
\citep{Woodward_1984} for solving the  hydrodynamic equations.  A
complete description of this method is not possible here and we will
only give short description (see \cite{Bryan_1995} for more
details). PPM  is a higher  order accurate version of  Godunov's
method with a third order  accurate piecewise parabolic monotonic
interpolation and a  non-linear Riemann  solver for shock
capturing. The  method is second  order accurate  in  space and  time
and explicitly  conserves energy mass flux and  momentum. 
It uses a dual energy formalism  which allows the
calculation of both  the thermal energy and the total energy of the
gas at each time step. This  ensures a correct internal energy 
and  the correct entropy  jump at  shock fronts and  the correct 
temperature   and pressure in   hypersonic flows.  
This represents a major difference  with  respect to SPH codes
which employ  an artificial viscosity to capture  shocks (see
\cite{Springel_2002}). \\ \indent  In addition to solving  the ideal
gas dynamics  \enzo also has  several cooling and heating
routines.  For the cooling both non-equilibrium and  equilibrium
cooling functions are available.

We will use here  a modified version of \textsc{Enzo}, which, uses the \gad
equilibrium chemistry solver  for hydrogen and helium with a uniform
ultra-violet(UV)  background based  on the models of
\cite{Haardt_1996}.

\subsubsection{The adaptive mesh refinement}
The  AMR ability of \enzo  introduces  finer and  finer  grids  into
areas   of  high density  thus allowing maximum resolution  where it
is actually needed   at a minimum computational  expense.   This
ability  to  dynamically  refine  the resolution  is essential  for
accurately tracking  the non-linear  collapse of   rapidly
evolving  density   fields.  For \lya
forest simulations the resolution in high density regions 
is less important and we  investigate here
simulations with and without AMR.  With AMR the hydrodynamical
equations are  initially solved on a uniform  grid, the  solutions are
monitored  and the  patches of  the initially uniform grid are
refined if certain refinement criteria are met.  Parent grids then
produce child grids. For cosmological studies a refinement factor of 2
is normally used.  This means that a child grid will have cells which have 
twice the spatial resolution of the parent grid. It is also worth noting that  
grids at the same level will have the same timestep but that this
timestep may be different for  grids 
at a different level of refinement. For \lya forest simulations it is
not obvious how important the use of the AMR option is. Most of the 
absorption especially at high redshift is by gas of moderate
(over)density.  Note that previous 
studies of the \lya forest with grid-based codes have generally not
used AMR methods (e.g. \citealt{Jena_2005}, \citealt{McDonald_2005b}). 
However as shown by \cite{Viel_2004b}  the few strong
absorption systems caused by dense regions contribute significantly to the flux
power spectrum at all scales.  We will investigate this  further in 
\S\ref{fluxpower}.

%Same thing!
%This means that a child grid will have a grid
%with twice as many grid points as the parent grid.

\subsection{\gad}

\gad (Springel 2005)\nocite{Springel_2005}, the updated version of
\textsc{Gadget-1} (Springel 2001)\nocite{Springel_2001} is a parallel 
\textsc{TreePM-SPH} code. On the scales relevant for the \lya forest
\gad in its TREEPM mode is similar to a PM  code with some extra resolution due to the
\textsc{Tree} part of the algorithm on small  and intermediate scales.  
The gravitational components of  \gad and \enzo are  therefore
somewhat similar on large scales most relevant for the \lya forest. 
The hydrodynamical components are, however,  very different. \\
\indent We  have used a version  of \gad which is similar
to the publicly available version as of August 2006.  The 
only exceptions are that we have used a \gad equilibrium cooling algorithm
supplied to us by the authors of \gad, and 
 that some of the simulations discussed in sections \S\ref{Sims} 
and \S\ref{fluxpower},  were run with a version of \gad optimised for speed 
for \lya forest simulations where gas  with an overdensity $>$ 
1000 and a temperature $<$ $10^5$ K is 
turned into collisionless star particles 
(see \citealt{Viel_2004} for more details). 

\subsubsection{The \gad gravity solver} 

We have used a version of \gad which employs a \textsc{TreePM}
algorithm to solve the gravitational equations (Xu 1995, Bode et al. 2000,
Bagla \& Ray 2003).
\nocite{Xu_1995} \nocite{Bode_2000}  \nocite{Bagla_2003}
 The \textsc{TreePM} algorithm
is a  hybrid of the tree \citep{Barnes_1986}  and particle mesh
methods (\citealt{Efstathiou_1985}, \citealt{Hockney_1988}).  
It utilises the best elements of both making  the
gravitational  force determination more  accurate and efficient.  The
potential is  split  into  two  components $\phi_k  = \phi^{\rm long}_k +
\phi^{\rm short}_k$, where

\begin{equation}
\phi^{\rm long}_k = \phi_k exp(-k^2 r^{2}_{s})
\end{equation}

and $  r_s $  is the  spatial scale of  the force  split which  is
usually set to  a little larger than the mesh  spacing. 
For the simulations 
performed here we used the \gad default value of $  r_s $ equal to 1.25 times the mesh spacing
and a mesh spacing equal to the cube root of the total number of dark matter particles. 
The long range
force is then computed using  mesh methods making the long range force
almost exact. The short range force is computed using a tree algorithm
which calculates the gravitational particle-particle forces on small
and  intermediate scales in an efficient manner.\\
\indent  In order to conserve the symplectic nature of the leap-frog 
time integration for the case of individual timesteps the Hamiltonian 
is separated into a kinetic part and a long range and short range potential. 
\gad then evolves all particles
using individual timesteps hence reducing the computational overhead that
would be associated with evolving all the particles using the minimum
allowed timestep. The splitting of the time integration is similar to what 
is done in the \textsc{TreePM} algorithm 
(see \citealt{Springel_2005} for more details).
\textsc{TreePM} codes offer an excellent compromise 
between speed and accuracy.

\subsubsection{The \gad Hydrodynamics solver}
The \gad hydrodynamics solver uses the SPH formalism. 
SPH can be thought of as a 
discretisation of a  fluid which is  then represented by particles.
Continuous fluid properties are then defined  using kernel
interpolation. The particles sample  the gas in  a Lagrangian  sense
thus  making SPH  methods very powerful   for   following   structure
formation   in   cosmological simulations. \\ \indent The
thermodynamic state of each fluid element can either  be defined in
terms  of its thermal energy  per unit mass, $u_i$, or  its entropy
per unit  mass, $s_i$. In  \gad the entropy per unit mass, $s_i$, is
used (see Springel \& Hernquist(2002)). The  code  conserves  
both energy  and entropy  even when fully adaptive
smoothing lengths are used. This  represents a major  change  in
methods  between \textsc{Gadget-1}  and  \gad  which  is investigated in
\cite{O'Shea_2005}. \\
\indent  A potential drawback of  SPH is the approximate way  in
which it captures shocks by use of an  artificial viscosity 
(see \cite{Springel_2005} for more details). 
\\  \indent The large differences in the methodology of the hydro-solvers 
make a comparison of  \lya  forest simulation with both codes very
interesting for a test of the sensitivity of these simulations to a particular numerical 
method. 
%%%%%%%%%%%%%%%%%%%%%%%%%%%%%%%%%%%%%%%%%%%%%%%%%%%%%%%%%%%%%%%%%%%%%%%%%%%
%%%%%%%%%%%%%%%%%FIGURE%%%%%%%%%%%%%%%%%%%%%%%%%%%%%%%%%%%%%%%%%%%%%%%
\begin{figure*}
  \centering           \begin{minipage}{175mm}          \begin{center}
  \psfig{figure=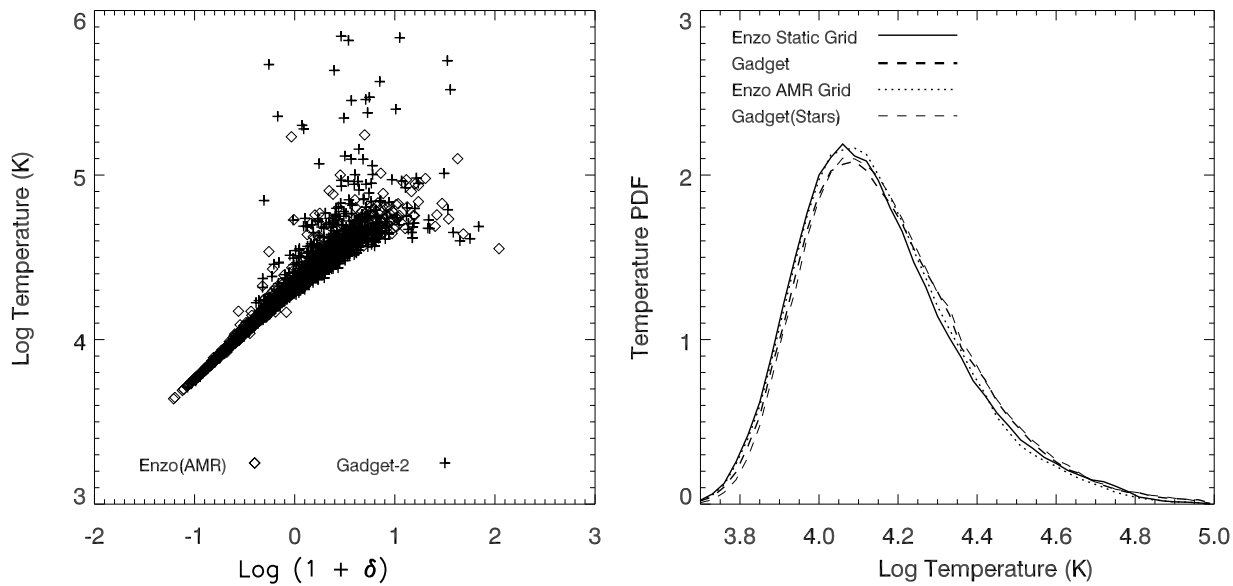,width=1.0\textwidth}
  \caption{\label{Temp_PDF}  {\it Left Panel:}   A  comparison  of  the Enzo(AMR only)
    and  \gad gas temperature over-density relation at $z = 3$ for the $(15,200)$ simulations. 
    The diamonds represent the \enzo simulation  while pluses represent the \gad
    simulation. 10000  values from each simulation are plotted.
    {\it Right Panel:} PDF of the volume-weighted gas temperature
    distribution for the $(15,200)$ simulations with \enzo with and without AMR and \gad 
    with and without simplified star formation. All the  results
    are at $z=3.0$.}

  \end{center} \end{minipage}
\end{figure*}

%%%%%%%%%%%%%%%%%%%%%%%%%%%%%%%%%%%%%%%%%%%%%%%%%%%%%%%%%%%%%%%%%%%%%%%%%%%%
\subsection{Simulation parameters}
\label{Sims}

We have performed simulations  with parameters of the concordance
cosmological model with  $\Omega_{\Lambda} = 0.74, \Omega_{\rm m} = 0.26,
\Omega_{\rm b} = 0.0463, \sigma_8 = 0.85, n = 0.95$ and  $h$ = 0.72. These
simulation parameters correspond  to the 'B2' model of
\cite{Viel_2004}.  Identical initial conditions were used for the
simulations with both codes.  \enzo was run with the implementation of  
the \gad equilibrium chemistry so that cooling
and heating were also treated in the same way.  
\cite{Theuns_1998b}, \cite{Viel_2004}, 
\cite{McDonald_2005b}, \cite{Jena_2005} and \cite{Bolton_2005} 
have investigated the effect of box size and 
resolution of hydrodynamical simulations on  a variety of statistics 
of the \lya forest flux distribution 
most importantly  the flux probability distribution, the effective optical depth 
and the flux power spectrum.  Unfortunately it is currently not
possible to run hydrodynamical simulations for which these statistics are fully
converged especially not for a large parameter space.  Convergence
and box size studies  are therefore essential for quantitative studies
of the \lya forest. Generally compromises have  to be made and the
application of corrections and an analysis of the
corresponding errors are necessary.  Our aim is here to investigate
how the uncertainties between codes employing different hydrodynamical
methods compare to other errors in a  quantitative analysis. 
We have thus ran simulations  with up to four different numbers of basic
resolution elements  (number  of  particles and  grid  cells,
respectively)  and for  three different  box  sizes.
The  \enzo simulations were  run  without  AMR and with up to 4 levels of grid
refinement. We have chosen a refinement level of 4 so as to match Spatial Dynamic
Range (SDR) of SPH calculations currently used in calculating the flux statistics of the
\lya forest. The simulation parameters are summarised in Table \ref{Gadget_sims}. 
Note that only the higher resolution simulations resolve the
Jeans mass well. We will come back to this later. \\
\indent  The SPH nature of the \gad simulations leads to a varying
resolution similar  to  that  of  an  AMR  code.  In  order to get a
feel for how  the resolution of the different simulations compare, the
last column  of Table \ref{Gadget_sims} gives the SDR.  
For the SPH simulation the SDR is calculated as $ {\rm
SDR}$ = L/$\epsilon$ where $\epsilon$  is the  gravitational softening
length  and is calculated before the simulation begins by dividing the
boxsize L by the number of particles along  one axis times some
constant factor. \\   
\indent  The  \enzo simulations were run with a
static grid and with AMR.  For \enzo the SDR is calculated
as $ {\rm SDR} = { N_{\rm root} \times 2^{l} }$, where $N_{\rm root}$,  is the
size of the root grid in 1D and $l$ is the refinement factor.  In the
static grid simulations the grids are  fixed  throughout  the
simulation without any  refinement. As
a  result the  spatial resolution  in high density regions  will  be
comparatively  poor in  these simulations. Most of the absorption  in
the \lya forest is, however,   produced by  regions of low or moderate
density. As we will see later the differences in the statistics of
the flux distribution  between simulations with and without AMR  are
therefore actually moderate. The static grid simulations would be comparable
in resolution to  a \gad simulation with  a softening length
equal to  the  mean  inter-particle spacing.  \\ 
\indent Unfortunately the improved resolution of the AMR  simulations comes 
at  the expense  of a significant increase in computational
time. For the AMR simulations  we set the maximum refinement level 
to 4 beyond which an artificial  pressure support is introduced to prevent further
collapse.  
We thereby experimented with the values of the parameters for the minimum 
pressure support and checked that the thermodyamic properties of 
the cells in question were not affected.
The mesh-refinement criterion was set to the
standard values of 4 for the dark matter (DM) and 8 for the baryons (see
\cite{O'Shea_2005} for more details).  This means that a grid will
refine when its DM density reaches a factor 4 greater than the root DM
density or when its baryonic density reaches a factor of 8 greater
than  the root baryonic density.   
To make contact with the simulations used in actual measurements of the matter
power spectrum from \lya forest data we also investigated  some of the
simulations used in \cite{Viel_2004}.
These simulations employ a simplified star formation criterion
that turns all  gas with an over-density, with
respect to the mean baryonic density, $> 1000$ and  temperature $<
10^5$ K into collisionless star  particles. This substantially
reduces the required computational time by eliminating the short
dynamic  time scales associated with high density gas 
and significantly
speeds up \lya forest simulations with \textsc{Gadget-2}.
The effect on the
statistics of the flux distribution has been shown to be small
(see \cite{Viel_2004b}) and we have labelled these simulations as \gad(stars).\\ 
Most of these simulations have very large particle numbers ($2\times
400^3$). Unfortunately for these simulations only a comparison with the
\enzo  static grid  simulations is  feasible with our limited computational
resources.
 %%%%%%%%%%%%%%%%%%%%%%%%%%%%%%%%%%%%%%%%%%%%%%%%%%%%%%%%%%%%%%%%%%%%%%%%%%%%%%%%%%%%
%%%%%%%%%%%%%%%%%%%%%%%%%%%%%%FIGURE%%%%%%%%%%%%%%%%%%%%%%%%%%%%%%%%%%%%%%%%%%%%%%%%%
\begin{figure*} 
  \centering \begin{minipage}{175mm} 
    \begin{center}
      \psfig{figure=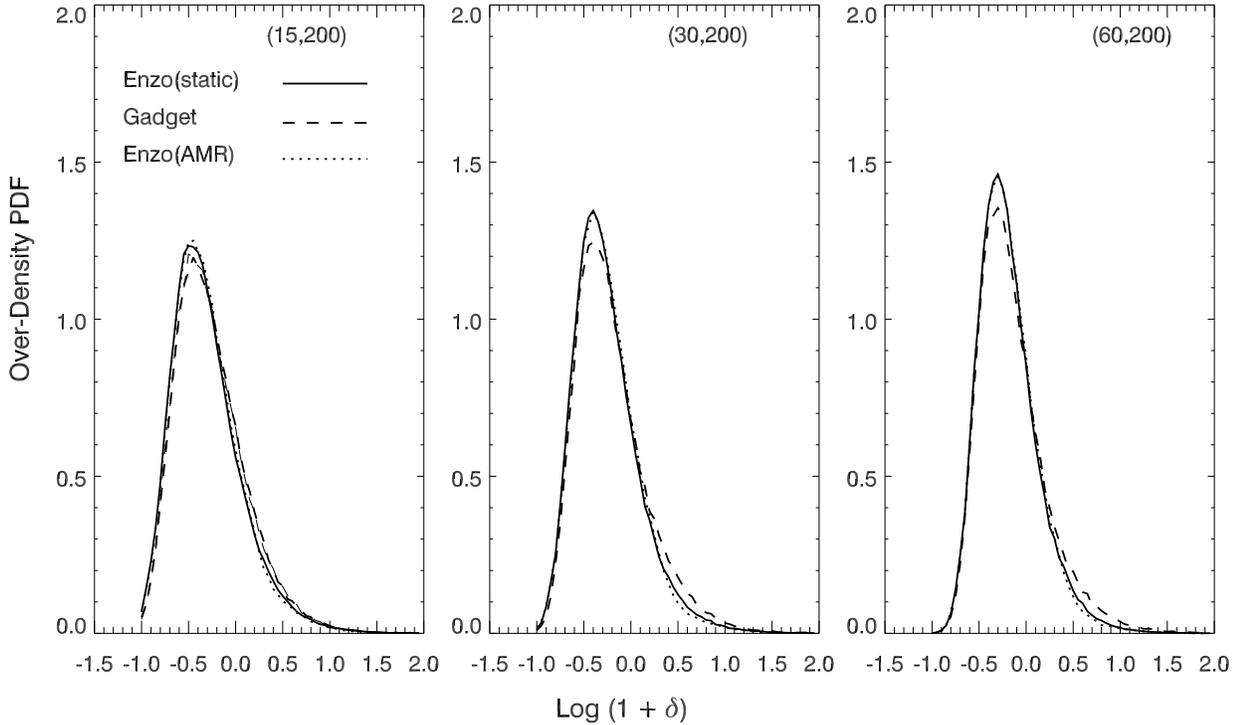,width=1.0\textwidth}
      
      \caption{\label{Gas_PDF} PDF of the volume-weighted gas density
         distribution for simulations with \enzo with and without AMR and \gad 
         with and without simplified star formation. The panels are for simulations
         with a boxsize of $15h^{-1}$,$30h^{-1}$ and $60h^{-1}$ Mpc
         (comoving)  from left to right. All the  results
        are at $z=3.0$.}
  \end{center} \end{minipage}
\end{figure*}
 %%%%%%%%%%%%%%%%%%%%%%%%%%%%%%%%%%%%%%%%%%%%%%%%%%%%%%%%%%%%%%%%%%%%%%%%%%%%%%%%%%%%%%
  We will concentrate our comparison on simulation outputs 
at $z=3$ the centre of the redshift range $2<z<4$ relevant for 
quantitative measurements of the matter power spectrum studies from 
\lya forest data,  but will briefly discuss simulations 
of the \lya flux power spectrum at $z=2$ and $z=4$
in \S\ref{fluxpower}.

%%%%%%%%%%%%%%%%%%%%%%%%%%%%%%%%%%%%%%%%%%%%%%%%%%%%%%%%%%%%%%%%%%%%%
%%%%%%%%%%%%%%%%%%%%%%%%%%%%%%FIGURE%%%%%%%%%%%%%%%%%%%%%%%%%%%%%%%%%
\begin{figure*} 
  \centering           \begin{minipage}{175mm}          \begin{center}
  \psfig{figure=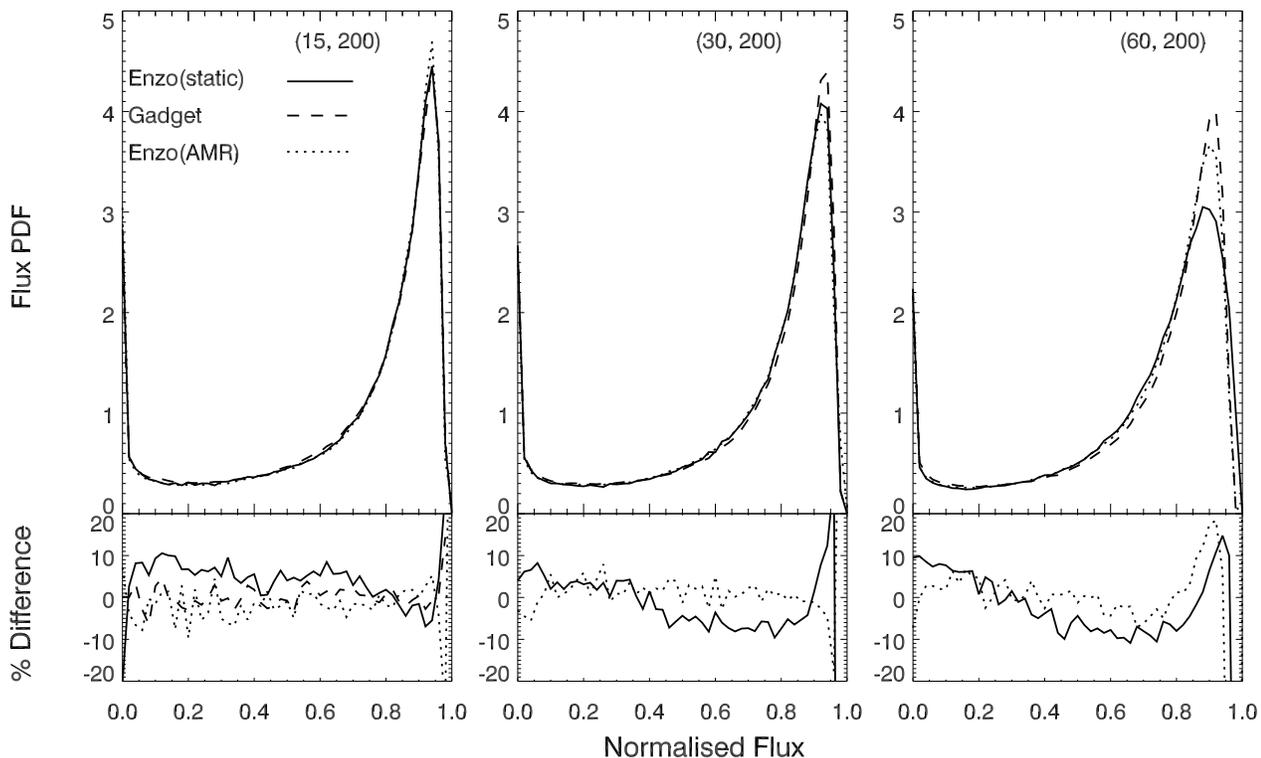,width=1.0\textwidth}
  %\caption{\label{Norm_flux} 
  \caption{\label{Flux_PDF} PDF of the flux 
    distribution for simulations with \enzo with and without AMR and \gad 
    with and without simplified star formation. The panels are for simulations
    with a boxsize of $15h^{-1}$,$30h^{-1}$ and $60h^{-1}$ Mpc (comoving) from left to right
    with $200^3$ particles/rootgridsize.
    The bottom panel shows the fractional  differences between  the \enzoamr simulation and the 
    \gad simulation (solid curve), the \enzoamr  and \enzost simulations(dotted curve) and between
    the \gad simulations with and without star formation (dashed curve in the left panel) 
   All the  results are at $z=3.0$.}
  \end{center} \end{minipage}
  
\end{figure*}
%%%%%%%%%%%%%%%%%%%%%%%%%%%%%%%%%%%%%%%%%%%%%%%%%%%%%%%%%%%%%%%%%%%%%%%

\section{Code Comparisons}
\label{comparison}
\subsection{The dark matter distribution}
We start with a  comparison of the DM distribution.  O'Shea et al. (2005) 
have  recently performed such a comparison and found moderate  differences but 
note that  we are interested  in a 
different application of the code than investigated in  O'Shea et al. (2005).  
The most relevant statistical property of the matter
distribution is the power spectrum $P(k) = < |\delta_k|^2 > $, 
where  $\delta_k$ is the Fourier transform of the density field.  
O'Shea found very good agreement at large scales  with deviations at 
small scales.  \\
\indent  The left panel of Figure \ref{DMPS} shows the DM power spectrum 
of the \enzoamr simulation in the form  $\Delta^2 (k) =  P(k) k^{3}$ 
for three simulations with a $200^3$ root grid  but different box
sizes at $z=3$. The bottom panel in the left panel of Figure \ref{DMPS} shows the
percentage difference  between simulations with and without AMR 
in the form [(\enzoamr - \enzost ) /
\enzoamr ]. As expected the AMR simulations show  more small scale power
as the particle mesh algorithm becomes more accurate  at small scales
due to mesh refinement.  The differences at large scales are very
small,  of the order of $1\%$.   In the right panel of Figure
\ref{DMPS}  we have plotted the fractional  difference
between the DM power spectrum of the \enzoamr and the \gad simulations
in the form [(\gad - \enzoamr)/\gad]. Similar to O'Shea we find
differences  of $\sim 1\%$ at large scales.\\  
The strong increase at small scales is due to the somewhat different
resolution limits of the  simulations compared. \cite{O'Shea_2005}
came to a similar conclusion and showed  that \enzo and \gad produce  
very similar results, even at small scales, when a low over-density 
threshold is used for the mesh-refinement criteria.
As pointed out by \cite{O'Shea_2005}, and verified by us, \gad has 
a higher force  resolution and hence more power at small scales due to
the better force resolution of the Tree algorithm  when simulations of
similar SDR are compared and standard parameters are used in both
simulations.  Note, however, that for simulations of the \lya forest 
the differences at  the relevant scales are very small.   
This makes a  meaningful comparison of the effect of the different 
hydrodynamics solvers on the statistics of the flux distribution 
-- the main aim of this paper -- possible.

\subsection{Properties of the gas distribution}

\subsubsection{Shock heating and the thermal state of the gas} 
\label{shocks}
As we described in \S\ref{Codes} at moderate  to  low over densities
($\delta \lesssim  5$) the  temperature  - density  relation
is well approximated by a power law. In the left panel of Figure 
\ref{Temp_PDF} we have plotted  the temperature - density 
relation for a simulation  with 2 $\times 200^3$ DM and gas 
particles/rootgridsize in a $15 h^{-1}$ Mpc box, hereafter labelled as a ($15,200$) simulation, 
for  \enzoamr and  \textsc{Gadget-2}.  10000 points  are plotted for each simulation.
This gives a feel for the level of shock heating produced by
each code and also  emphasises that the vast majority of the gas lies
very close to a line representing the power law approximation. The
temperature  is volume-weighted in both cases. We have chosen
volume-weighted temperatures as the flux statistics of the 
\lya forest are volume-weighted. Temperatures and densities were 
calculated in the same way  as for the mock absorption spectra.
Note that the differences for the 
mass-weighted temperatures were somewhat smaller.   \\   
%The simulations with \gad show slightly more shock heating 
%but 
Overall the  agreement between the two codes is remarkable 
given the very different ways in which both codes treat  shocks.
As mentioned previously \gad  uses a conservative  entropy
formalism  to treat  shocks while \enzo uses a  non-linear Riemann
solver. In principle  since \gad employs an artificial
viscosity  to capture shocks \enzo  should  resolve  shocks more
accurately and one may have expected that weak shocks may occur in 
low density regions which \gad does not capture properly.
This appears not to be the case. The amount of shock  heated 
gas and its temperature distribution is very similar. This can be seen 
more clearly in the right panel of Figure \ref{Temp_PDF} where we have plotted  the volume weighted
probability distribution function (PDF) of the temperature for 
simulations with \enzo with and without AMR and for \gad with and without the 
simplified star formation criterion.  The differences are of order 10\% 
and can be at least partially attributed to differences in the PDF 
of the density which we will discuss in the next section.

\subsubsection{The probability distribution function of the gas density}

Figure  \ref{Gas_PDF} shows the volume-weighted  PDF of the gas 
distribution for simulations with a  range of box sizes. 
The agreement between the \enzo simulations with and without 
AMR in this linear volume-weighted plot which emphasises gas around 
the mean density is very good. The differences between the simulations 
with \gad and \enzo are somewhat larger,  of order 10\%. Note, however, that this is 
smaller than the  differences due to changes 
in box size and resolution. Overall the agreement is again 
very good.

%The mass resolution of the above
%simulations is also worth considering. The mass resolution of the \gad simulations and 
%the \enzost simulations are $7.528 \times 10^6 M_{\sun} $, $6.023 \times 10^7 M_{\sun}$ 
%and $4.818 \times 10^8 M_{\sun} $ for the (15, 200), (30, 200) and (60, 200) simulations
%respectively. In order to resolve the Jeans mass in these simulations a mass resolution of 
%at least $10^8 M_{\sun}$ is needed(see \citealt{Schaye_2001}). The mass resolution of the
%\enzoamr simulations is adaptive but it only adapts in high density regions so regions
%of low to moderate over-density suffer from the same problems as both the \enzost grid and \gad. 
%The mass resolution effect is an important consideration 
%and should be kept in mind by the reader especially when considering the effect of
%resolution on the flux PDF and flux power spectrum.} \\
\indent As demonstrated in the previous sections, 
a SPH code and a grid-based
code differ in  their resolution  properties and it  is not trivial
to run simulations with the ``same  resolution'' because of
differences in  force resolution and the way the resolution is
distributed between regions of different densities. The (small)
differences between simulations with \enzo and \gad 
are thus not surprising. 

\subsubsection{The probability distribution of the flux}

We have computed the flux distribution for  1000 random 
lines of sight through the simulation box. The optical depth 
has been rescaled in the standard  way to match the observed 
effective optical depth at $z=3$ as given by
\cite{Schaye_2003}, $\tau_{\rm eff} =  0.363$.  
Figure \ref{Flux_PDF} shows the corresponding 
probability distribution of the flux for simulations with \enzo 
with and without AMR and \gad with and without simplified star 
formation. Overall the flux distributions are very similar. 
Typical differences between the simulations with \enzo and \gad are 
5-10\%. The differences are again smaller than those due to changes 
of box size and resolution. The differences between the \gad
simulations with and without star formation are less than 3\%.  

%%%%%%%%%%%%%%%%%%%%%%%%%%%%%%%%%%%%%%%%%%%%%%%%%%%%%%%%%%%%%%%%%%%%%%%%
%%%%%%%%%%%%%%%%%%%%%%%%%%%%FIGURE%%%%%%%%%%%%%%%%%%%%%%%%%%%%%%%%%%%%%%
\begin{figure}
 
  \psfig{figure=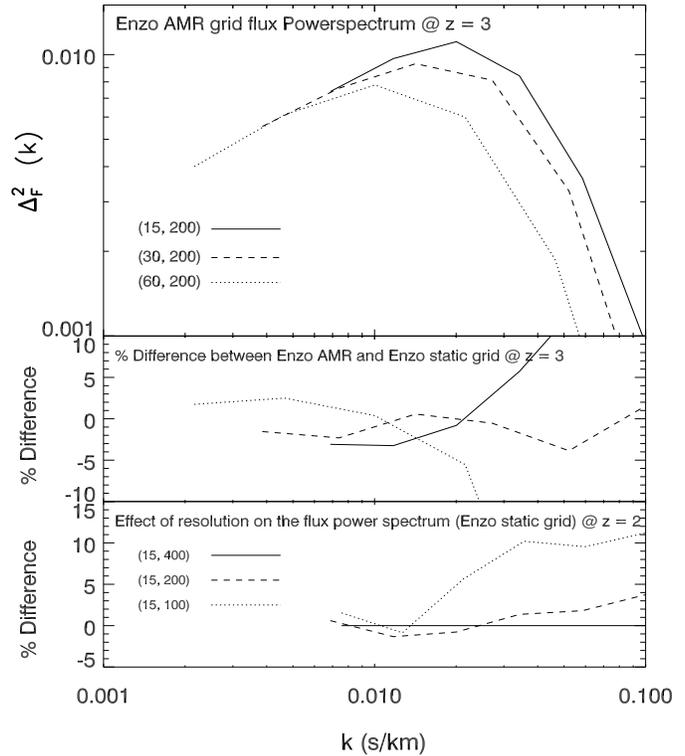,height=110mm, width=90mm}

  \caption{\label{static_amr_fluxpower} The top panel shows the 
    flux power spectrum for simulations with \enzoamr
    with a boxsize of $15h^{-1}$,$30h^{-1}$,$60h^{-1}$ Mpc comoving. 
    The middle panel shows fractional differences 
    between  the \enzo simulation with and without AMR [\enzoamr -
    \enzost )/\enzoamr ]. The top and middle panel are for $z=3$ 
    and the line styles are identical. The bottom panel shows the 
    effect of resolution on the flux power spectrum at $z=2$. 
    The differences are relative to the highest resolution run e.g. 
    ((15, 400) - (15,200)) / (15,400).
    }
 
\end{figure}
%%%%%%%%%%%%%%%%%%%%%%%%%%%%%%%%%%%%%%%%%%%%%%%%%%%%%%%%%%%%%%%%%%%%%%%%

\subsubsection{The flux power spectrum}
\label{fluxpower}
In  Figure \ref{static_amr_fluxpower}, we show the flux power spectrum  for the
\enzo AMR simulations for different boxsizes at $z=3$.  The middle  panel shows 
the fractional difference between \enzo simulations with and without AMR. 
The solid curve is for  simulations with a boxsize  15 $h^{-1}$
Mpc, the dashed curve is for a boxsize of  30 $h^{-1}$  Mpc box and the
dotted curve is for a box size of  60 $h^{-1}$ Mpc. 
At large scales the differences are less than 4\%. 
At the resolution
limit the  differences increase as expected.  The effect of the AMR is 
most significant  at small scales. Note that the force resolution limit 
(twice the cell length) is about an order of magnitude 
off the graph on the right hand side. 
The bottom panel of Figure \ref{static_amr_fluxpower} demonstrates the
convergence of the flux power spectrum by comparing the flux power
spectrum for the static grid simulations with a box size of 15
$h^{-1}$ Mpc at $z=2$.  
The differences between the (15,200) and (15,400) simulations are less
than 2 percent  on the relevant scales suggesting that a resolution of 150$h^{-1}$ kpc
(comoving) is required to reach convergence.  This is in good
agreement with the results by \cite{Viel_2004} for the \gad 
simulations also used here. Note, however,  that a similar 
comparison by Jena et al. (2005) for static grid \enzo simulations
with the same resolution as shown in the bottom panel (their Figure 7)
showed significantly larger differences. A discrepant result for which
we do not have an explanation. 

We have also investigated  how the level of AMR refinement effects the
\lya flux power spectrum. The results for simulations with refinement
level 2 lie in between those with refinement level 4 shown 
in Figure \ref{static_amr_fluxpower} and the static grid simulations.   
This suggests that a relatively high refinement level may  be
needed to correctly account for the effect of strong absorption
systems on the flux power spectrum caused by high density gas 
which \cite{Viel_2004b} have shown to extend to large scales.

%%%%%%%%%%%%%%%%%%%%%%%%%%%%%%%%%%%%%%%%%%%%%%%%%%%%%%%%%%%%%%%%%
%%%%%%%%%%%%%%%%%%%%%%%%%%FIGURE%%%%%%%%%%%%%%%%%%%%%%%%%%%%%%%%%%% 
\begin{figure}
 
  \psfig{figure=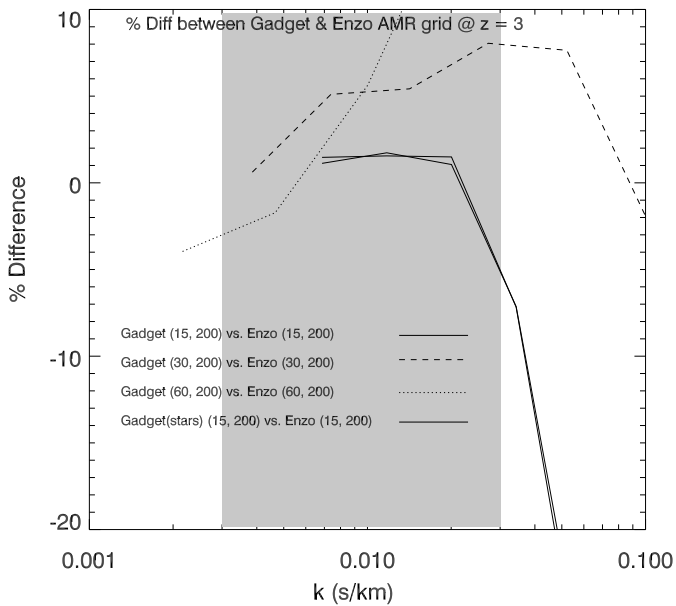,width=0.48\textwidth}
  \\
\caption{  \label{amr_fluxpower_ratio} The fractional difference 
   of the flux power spectrum of simulations with \enzoamr and \gad 
   for different box sizes [(\gad - \textsc{Enzo}(AMR))/\gad]. Also shown is the difference 
   for \gad simulations with and without star formation. The shaded 
  region indicates the range of wave numbers used by Viel et al. (2004) 
  to infer the linear dark matter power spectrum.} 
\end{figure}

%%%%%%%%%%%%%%%%%%%%%%%%%%%%%%%%%%%%%%%%%%%%%%%%%%%%%%%%%%%%%%%%%%%%%%

\begin{figure}
 
  \psfig{figure=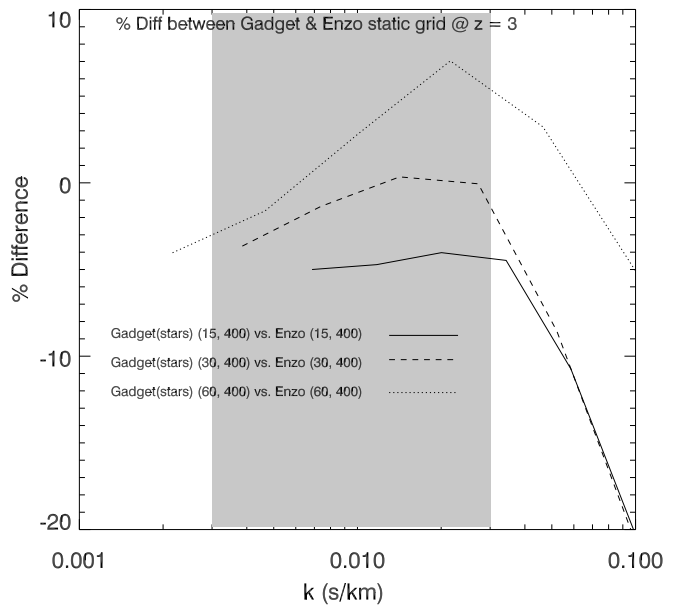 ,width=0.48\textwidth}
  \\
\caption{\label{amr_fluxpower_ratio_400} The fractional difference 
   of the flux power spectrum of simulations with \enzost and \gad 
   for $400^3$ gas particle/rootgridsize simulations with different box sizes. 
   The difference is of the form [(\gad - \textsc{Enzo(static)})/\gad]. All Gadget 
   simulations are in this case run with the simplified star formation algorithm. 
   The shaded region indicates the range of wave numbers used by 
   Viel et al. (2004) to infer the linear dark matter power spectrum. } 
\end{figure}

In Figure \ref{amr_fluxpower_ratio} we show the fractional difference 
of the flux power spectrum of the \gad and \enzo simulations for the $200^3$ 
simulations. Apart from the smallest scales in the lowest resolution
simulation the differences are about 5\%. The differences are scale
dependent and appear to decrease with increased resolution. Unfortunately we did not have the
computational resources available to run \enzoamr simulation with a
root-grid size of $400^3$.  In Figure \ref{amr_fluxpower_ratio_400} 
we therefore show the differences between $400^3$ \gad and \enzost
simulations. The differences are again about 5\%. The reader should 
thereby keep in mind the differences of up to 4\% between the \enzoamr and \enzost
simulations. Note that Viel et al. (2004) found the 
(60,400) simulations to be the best compromise between box size and resolution in their
measurement of the matter power spectrum from \lya forest data.  
We have also looked into the differences at $z=2$ and $z=4$ and found
the differences to be redshift dependent. At $z=4$ the differences are
similar than those at $z=3$ while at $z=2$ they are somewhat larger.  
By investigating the $z=3$ simulations with different effective
optical depth we verified that the change of the differences 
of  the simulated flux power spectra with redshift is
mainly but not only due to the strong evolution  of the effective
optical depth.  Also worth noting is the excellent agreement
between the  \gad simulation with (thin solid curve)  and without star
formation (thick solid curve) shown in Figure
\ref{amr_fluxpower_ratio}. 
\\

%%%%%%%%%%%%%%%%%%%%%%%%%%%%%%%%%%%%%%%%%%%%%%%%%%%%%%%%%%%%%%%%%
%%%%%%%%%%%%%%%%%%%%%%%%%%FIGURE%%%%%%%%%%%%%%%%%%%%%%%%%%%%%%%%%%% 

%%%%%%%%%%%%%%%%%%%%%%%%%%%%%%%%%%%%%%%%%%%%%%%%%%%%%%%%%%%%%%%%%%%%%%%%%%%%%%

\subsection{CPU time requirements}
\label{time}

\begin{figure}
  
  \psfig{figure=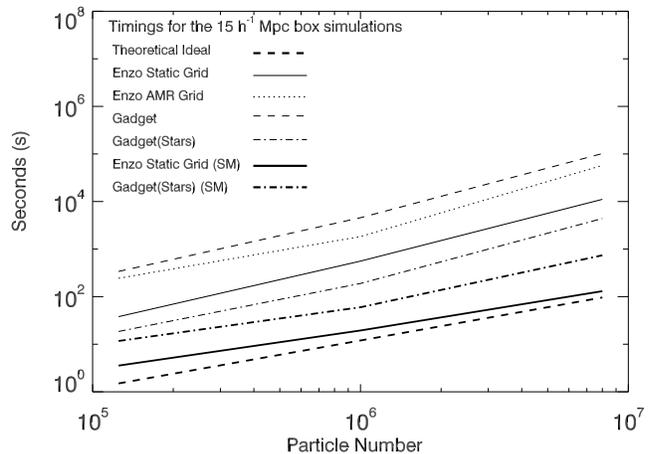,width=0.48\textwidth}
  \caption{\label{times} A comparison of the CPU time required for 
   the simulation to run through a fixed redshift interval. Simulation
   parameters are as described in the text and annotated on the
   plot. The thin lines are for a SUN distributed memory cluster and 
   the thick lines are for a SGI shared memory (SM) machine. Only relative
   values for the same architecture should be considered. 
   Note the reversal in relative speed between \gad simulations with
   star formation  and \enzo static grid simulation between the two
   architectures. The thick dashed line shows a linear scaling of CPU
   time with the number of particles for  comparison. 
}
    
\end{figure}

We have performed a series of timing tests for both \enzo and \gad
for a selection of the simulations used in this study.  For the timings,
simulations were carried out on the distributed memory 
Sun Cluster at the Institute of Astronomy in Cambridge and \cosmos a shared memory
machine located at the Department of Applied Mathematics and Theoretical Physics (DAMTP)
in Cambridge. 
On the distributed memory machine the timing tests were
performed on 32 single node 500 Mhz processors with 2 Gbytes of memory and 
a 100 Mbit ethernet connect. The latency of the connection is approximately
300 microseconds. Although these processors 
are relatively slow by today's standards they should still provide 
an illustration  of the relative performance of the codes in different configurations. 
The tests were performed for simulations with a boxsize of 15 $h^{-1}$ Mpc 
comoving for particle (rootgridsize) numbers $50^3$, $100^3$, $200^3$. 
The results are shown in Figure \ref{times}. We ran 
the \enzo simulations for 10 time steps starting at $z =
3.5$. For \gad we ran  the simulation from $z = 3.5$ 
to the same redshift that \enzo had reached. 
The \enzoamr simulations are about a factor 1.5-2 faster than the \gad 
simulations without star formation. Turning the AMR off leads to a speed-up 
of a factor five for  the \enzo simulations.  
The \gad simulations without  star 
formation  spends most of its time calculating the
hydrodynamics of the very high density gas, which for the \lya forest is 
not necessary. Turning on the star formation in \gad leads to a speed-up by a
factor 30. As we have demonstrated here (see also \cite{Viel_2004} ) the effect 
of turning on star formation in \gad on the statistics  of the flux 
distribution is very small. \\
\indent \cosmos is an SGI Altix 3700 with 152 Itanium2 (Madison) processors and  
152GB of globally shared main memory. Each Madison processor has a
clock speed of 1.3GHz, a 3MB L3 cache and a peak performance of 5.2 Gflops.
The system is built from 76 dual-processor nodes, each with 2GB of
local shared memory. These are linked by the SGI NUMAflexIII
interconnects, which provides a high speed (3.2GB/sec bi-directional),
low latency (sub-microsecond) network with a dual-plane, fat tree
topology, connecting all processors with each other and with a single,
globally shared and cache-coherent 152GB memory subsystem. We have
only tested the two faster versions of the codes on the shared memory machine.  
The results are shown in Figure \ref{times}. Note
the reversal in relative speed between \gad simulations with
star formation  and \enzo static grid simulation between the two
architectures. Obviously \enzo benefits more strongly from the 
shared memory  architecture (see  \cite{O'Shea_2005} for a similar result).  
The \enzost simulation run about a factor three faster on the shared 
memory machine than the  \gad simulations with star formation. 
Reducing the refinement level in the AMR simulation may thus offer a good
compromise between accuracy and speed for \lya forest simulations with
\enzo.  

\section{Discussion and conclusions}

We have performed a detailed comparison of \lya forest simulations 
with \gad, a \textsc{TreePM-SPH} code, and \enzo a Eulerian AMR code
in order to asses the numerical uncertainties due to a particular 
numerical implementation of solving the hydrodynamical equations. 
The codes are similar with respect to the way in which 
they compute the gravitational forces at large scales 
but differ in the way they calculate gravitational forces on small
scales; the codes use a Tree-PM and 
PM N-body algorithm, respectively. Their main differences lie,
however,  in the way in which they solve the gas  hydrodynamics. 
\gad discretises mass using SPH methods while \enzo  discretises 
space using adaptive meshes. The main results are as follows.

\begin{itemize}

\item{The differences in the dark matter power spectrum between simulations   
      with \enzo and \gad on scales relevant for measurements of the 
      matter power spectrum from \lya forest data are about 2\%
      for an appropriate choice of box size and resolution.}

\item{The temperature density relation of simulations with \enzo
      and \gad differ very little. 
      %The simulations with \gad show slightly more shock heating. 
      The PDF of the volume weighted 
      temperature differ by $\sim 10$ \% probably mainly due to 
      differences in the PDF of the gas density which are of the same 
      order and at least partially caused by a slight  mismatch in 
      resolution.}

\item{The PDF  of the flux distribution of simulations with 
      \enzo and \gad agree very well. Typical differences  
      are $\sim 5-10$\% probably again mainly due to a slight 
      mismatch of the resolution.}

\item{The differences of the flux power  spectrum of simulations   
      with \enzo and \gad on scales relevant for measurements of the 
      matter power spectrum from \lya forest data are about 5\%
      for an appropriate choice of box size and resolution and simulations
      which fully resolve the Jeans mass. For simulations of lower resolution 
      but larger boxsize the difference increase up to  $\sim 10$\%.
      Note that the differences are scale and redshift dependent. } 

\end{itemize}

%%%%%%%%%%%%%%%%%%%%%%%%%%%%%%%%%%%%%%%%%%%%%%%%%%%%%%%%%%%%%%%%%
%%%%%%%%%%%%%%%%%%%%%%%%%%FIGURE%%%%%%%%%%%%%%%%%%%%%%%%%%%%%%%%%%% 
%%%%%%%%%%%%%%%%%%%%%%%%%%%%%%%%%%%%%%%%%%%%%%%%%%%%%%%%%%%%%%%%%

Overall the \lya forest simulations with \enzo and \gad agree astonishingly 
well. The choice of method for solving the hydrodynamical simulations 
appears to affect the gas distribution and its thermal state very little. 
It is also reassuring that two different implementations for solving the 
gravitational equations agree well.  The corresponding uncertainties 
should contribute to the overall  error budget of measurements of the 
matter power spectrum from \lya forest data at the level of 3\%. 
The total error in current measurement is significantly larger  and they should 
thus not be important. The main numerical uncertainties
are  instead due to a lack of sufficient dynamic range which typically makes  
correction of 5\% for boxsize and resolution necessary. This will obviously 
improve as computational resources  become more powerful. In practical 
terms memory requirements of simulations with \enzo without AMR and
\gad are similar. \enzo without AMR offers the highest speed but 
requires somewhat larger corrections.       
Our results suggest that if sufficient computational resources 
are available and sufficient care is employed the accuracy of numerical simulations  
should  not yet be a limiting factor in improving the accuracy of
measurements of the matter power spectrum from \lya forest data.

\section*{Acknowledgements}
The simulations were run on the \cosmos (SGI Altix 3700) supercomputer
at DAMTP in Cambridge and on the Sun Sparc-based Throughput Engine 
at the Institute of  Astronomy in Cambridge. \cosmos is a UK-CCC facility
which is supported  by HEFCE and PPARC. We are grateful  to   Brian
O'Shea, Darren Reed and Volker Springel for  useful discussions and 
would like to thank the referee for a detailed report. This research 
was supported in part by PPARC and the  National Science Foundation 
under Grant No. PHY99-07949. 

\bibliographystyle{mn2e}

%\bibliography{/home/regan/MNRAS_30Aug/EnzoGadgetPaper/mybib}
%\bibliography{/data/cass55b/regan/MNRAS_30Aug/EnzoGadgetPaper/mybib}
\end{document}